\documentclass[journal=aamick,manuscript=article]{achemso}

\usepackage[version=3]{mhchem} 

\author{Levi C. Felix}
\affiliation[State University of Campinas]
{Applied Physics Department, State University of Campinas, Campinas, SP, 13083-970, Brazil}
\alsoaffiliation[State University of Campinas]
{Center for Computational Engineering and Sciences, State University of Campinas, Campinas, SP, 13083-970, Brazil}
\author{Vladimir Gaál}
\affiliation[State University of Campinas]
{Applied Physics Department, State University of Campinas, Campinas, SP, 13083-970, Brazil}
\author{Cristiano F. Woellner}
\affiliation[Federal University of Paraná]
{Physics Department, Federal University of Paraná, UFPR, Curitiba, PR, 81531-980, Brazil}
\author{Varlei Rodrigues}
\affiliation[State University of Campinas]
{Applied Physics Department, State University of Campinas, Campinas, SP, 13083-970, Brazil}
\author{Douglas S. Galvao}
\email{galvao@ifi.unicamp.br}
\affiliation[State University of Campinas]
{Applied Physics Department, State University of Campinas, Campinas, SP, 13083-970, Brazil}
\alsoaffiliation[State University of Campinas]
{Center for Computational Engineering and Sciences, State University of Campinas, Campinas, SP, 13083-970, Brazil}

\title{Mechanical Properties of Diamond Schwarzites: From Molecular Dynamics Simulations to 3D Printing}


\begin{document}


\begin{abstract}
   Schwarzites are porous crystalline structures with Gaussian negative curvature. In this work, we investigated the mechanical behavior and energy absorption properties of two carbon-based diamond schwarzites (D688 and D8bal). We carried out fully atomistic molecular dynamics (MD) simulations. The optimized MD atomic models were used to generate macro-scale models for 3D-printing  (PolyLactic Acid (PLA) polymer filaments) through Fused Deposition Modelling (FDM). Mechanical properties under uniaxial compression were investigated for both the atomic models and the 3D-printed ones. Mechanical testings were performed on the 3D-printed schwarzites where the deformation mechanisms were found to be similar to those observed in MD simulations. These results are suggestive of a scale-independent mechanical behavior that is dominated by structural topology. The structures exhibit high specific energy absorption and crush force efficiency $\sim 0.8$, which suggest that the 3D printed diamond schwarzites are good candidates as energy-absorbing materials.
   
\end{abstract}

\section{Introduction}

First studied by Hermann Schwarz in 1890, Triply-Periodic Minimal Surfaces (TPMS) possess locally minimized surface area under the constraint of periodic boundary conditions \cite{schwarz_1890}. There are several families of TPMS possessing different topologies satisfying such conditions. Examples of TPMS families proposed by Schwarz include Primitive (P) and Diamond (D) surfaces. Several other families were obtained almost a century later by Alan Shoen \cite{schoen_1970}, such as Gyroid (G) and I-graph Wrapped Package-graph (I-WP). Despite being initially a purely mathematical investigation, the subject of TPMS has been explored in materials science as an optimized surface topology with promising mechanical properties \cite{alketan_2019,sychov_2018,park_2019,restrepo_2017,ali_2017,se_2018,chen_2019}. Interestingly, they have also been observed in nature\cite{han_2018}. Examples include butterfly wings\cite{michielsen_2008,pouya_2016}, sea urchins\cite{lai_2007}, and soap films draped onto wire skeletons\cite{schoen_1970}. Also, G surfaces were obtained as a result of the self-assembling of block copolymers \cite{matsen_1996,bates_1999,feng_2019}.

Regarding structural applications, conventional fabrication methods at the nanoscale, such as Chemical Vapour Deposition (CVD), have serious difficulties to be employed due to the complex geometry of TPMS. However, additive manufacturing techniques, commonly referred to as 3D printing, have shown to be capable of reproducing structures with arbitrary geometries. Different 3D printing methods have been employed on the fabrication of TPMS structures such as Selective Laser Melting\cite{se_2018,yang_2018slm}, Selective Laser Sintering (SLS)\cite{maskery_2018}, Powder Bed Fusion (PDF)\cite{yuan_2019}, STereoLithograpy (STL)\cite{blanquer_2017,yu_2019,zheng_2018} and Fused Deposition Modelling (FDM)\cite{sychov_2018,yan_2019}. 

Motivated by carbon nanomaterials such as nanotubes and fullerenes, Mackay and Terrones\cite{mackay_1991} proposed a class of curved atomic structures called schwarzites (after Schwarz), where their geometry resembles the shape of TPMS. Previous works have shown that structural stability\cite{terrones_1992,vanderbilt_1992,townsend_1992,lenosky_1992,okeeffe_1992,terrones_1997,benedek_2003}, mechanical\cite{park_2010,miller_2016,felix_2019}, electronic\cite{phillips_1992,huang_1993,gaito_1998,valencia_2003,terrones_2003,weng_2015,owens_2016} and thermal properties\cite{pereira_2013,zhang_2017,zhang_2018} can be correlated to their negative Gaussian curvature.

The role of the surface curvature on the mechanical properties of schwarzites and other carbon nanomaterials has been highlighted in recent works that combine Molecular Dynamics (MD) simulations with 3D printing techniques\cite{qin_2017,sajadi_2018,sajadi_2019}. The optimized atomistic structures from MD simulations are translated into surfaces that generate 3D-printed macroscopic structures. The obtained structures are used to study and contrast the mechanical behavior at different length scales. This approach has been used to study the mechanical properties of P and G schwarzites\cite{sajadi_2018}. Thus, it is important to extend previous studies to include other schwarzite families. 

In this work, we investigated the role of geometry/topology on the mechanical and energy-absorption properties of atomic and 3D-printed schwarzites belonging to the Diamond (D) family. In Figure \ref{fig:structures}, we present the studied schwarzite atomistic models, which are pure carbon porous crystals. In Figure \ref{fig:surfaces} we present the surface models and the corresponding 3D-printed ones (Poly Lactic Acid (PLA), see details in Materials and Methods section). The (atomistic and 3D printed) schwarzite structures (D688 and D8bal) were uniaxially compressed beyond the elastic regime until the mechanical failure (fracture). The elastic properties, such as Young's modulus ($E$) and Poisson's ratio ($\nu$), were obtained from linear regime of the stress-strain curves. Different length scale models show qualitative agreement in the deformation mechanisms, specially on predicting a near-zero Poisson's ratio for one of the structures (D8bal). Also, due to the cellular character of schwarzites we calculated their energy-absorption characteristics and contrasted against other works on the literature regarding TPMS structures.

\section{Materials and Methods}

\begin{figure}[ht]
    \centering
    \includegraphics[scale=0.13]{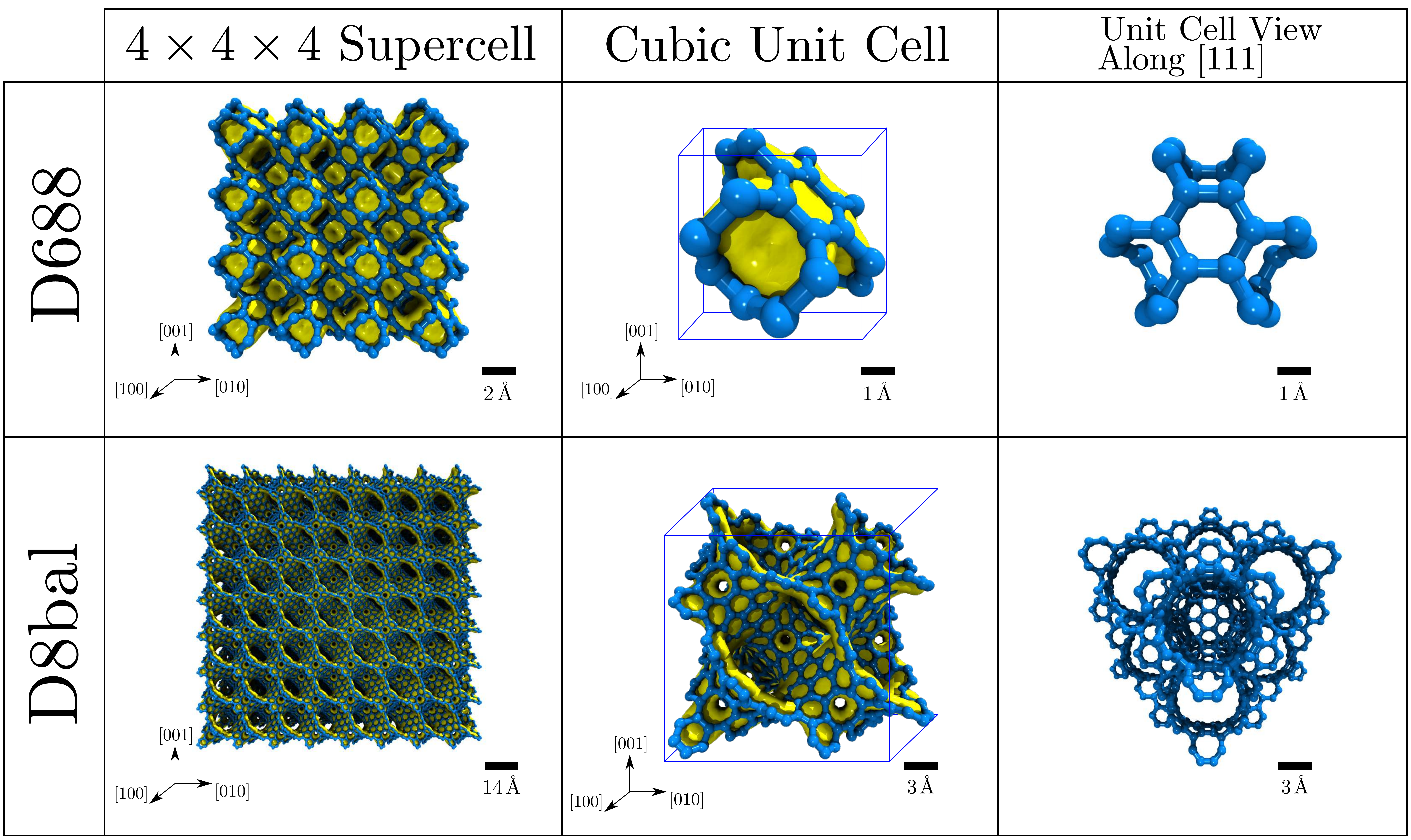}
    \caption{D688 and D8bal schwarzite structures studied in this work. (a) 4x4x4 supercell; (b) cubic unit cell, and; (c) view of the cubic unit cell along the [111] direction.}
    \label{fig:structures}
\end{figure}

\begin{figure}[ht]
    \centering
    \includegraphics[scale=0.2]{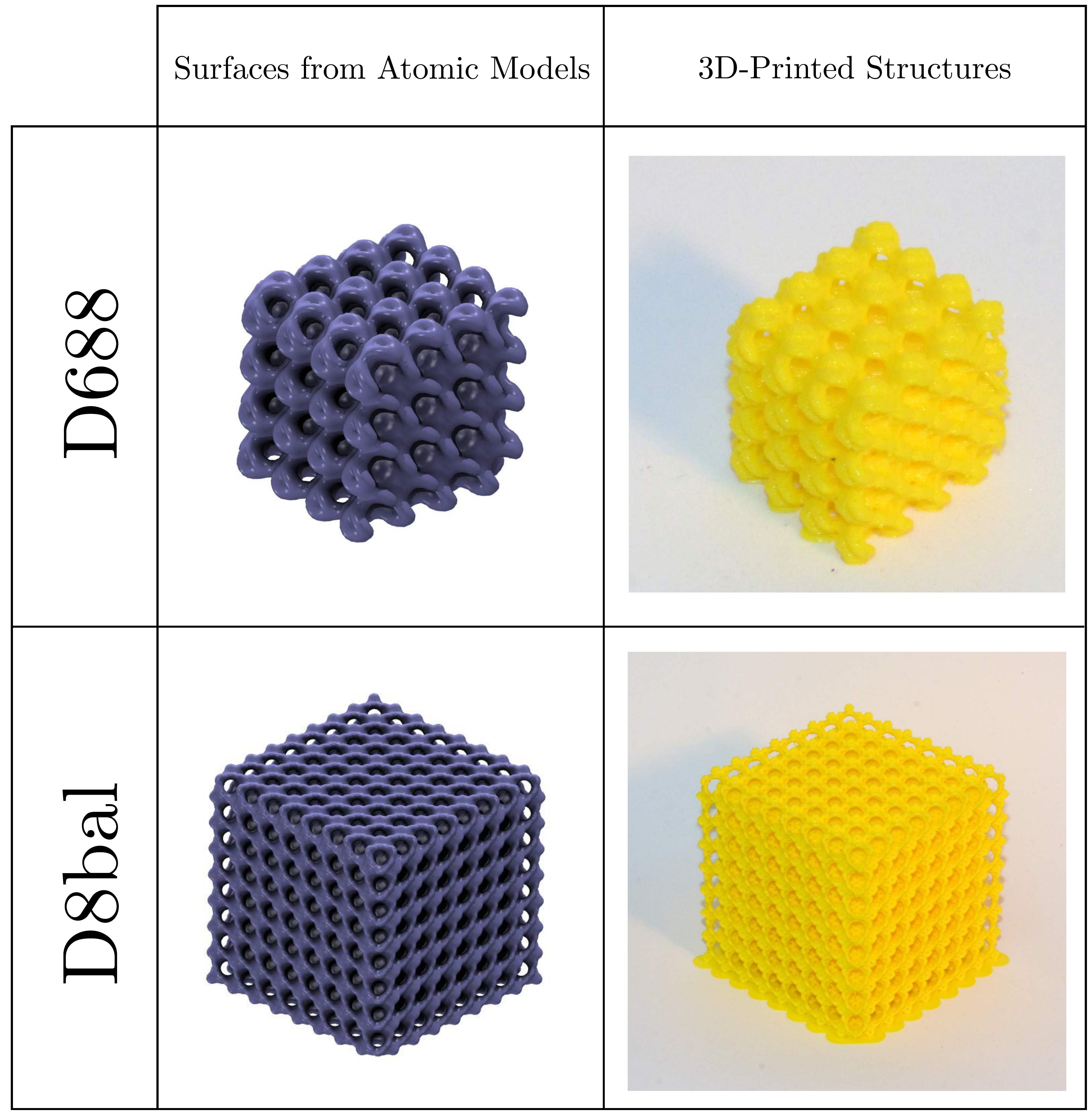}
    \caption{Surfaces from atomic models (left) used as input geometries to fabricate the macroscopic 3D-printed structures (right).}
    \label{fig:surfaces}
\end{figure}

The schwarzites studied in this work are shown in Figures \ref{fig:structures} and \ref{fig:surfaces}. We chose two representative structures from the D family, they differ mainly from the ratio of hexagonal to octagonal rings, which we will refer to as the "flatness" of the structures. The "flatness" is associated with the minimal number of hexagons separating two closest octagons. The structure denoted as D688 has no hexagons separating two consecutive octagons, whereas D8bal has three. Regarding the nomeclatures of both structures, D688 belongs to the D family where each six-membered ring is shared between two eight-membered rings, thus -688 suffix. The curvature of D8bal structure is consequence of the presence of eight-membered rings. The surface divides the structural space into two sub-domains of equal (or balanced) volumes, thus the -8bal suffix. For the mechanical and energy-absorbing analyses we considered 4$\times$4$\times$4 supercells, as indicated in Figures \ref{fig:structures} and \ref{fig:surfaces}. The use of these large supercells precludes spurious size effects. for the convergence of the mechanical properties with system size. The structural properties of the structures for both MD simulations and 3D-printed structures are listed in Table \ref{tab:structural}.

\begin{table}[ht]
    \centering
    \begin{tabular}{|c|c|c|c|c|c|c|c|}
    \hline
        \textbf{Structure} & $\boldsymbol{n}$ & $\boldsymbol{N}$ & $\boldsymbol{a}$ \textbf{[\AA]} & $\boldsymbol{l_{MD}}$ \textbf{[\AA]} & $\boldsymbol{\rho_{MD}}$ $\boldsymbol{[g/cm^3]}$  & $\boldsymbol{l_{3D}}$ \textbf{[mm]} & $\boldsymbol{\rho_{3D}}$ $\boldsymbol{[g/cm^3]}$  \\
        \hline
        D688 & 24 & 1536 & 6.2 & 22.9  & 2.6 & 23.5 & 0.33 \\
        \hline
        D8bal & 768 & 49152 & 23.7 & 91.1  & 1.3 & 90.4 & 0.19\\
         \hline
    \end{tabular}
    \caption{Structural properties of D688 and D8bal: number of atoms per cubic unit cell $n$ and in the supercell $N$, cube size for MD simulations ($l_{MD}$) and 3D-printing ($l_{3D}$) and their mass densities $\rho_{MD}$ and $\rho_{3D}$, respectively.}
    \label{tab:structural}
\end{table}

\subsection{Molecular Dynamics Simulations}

\begin{figure}[ht]
    \centering
    \includegraphics[scale=0.35]{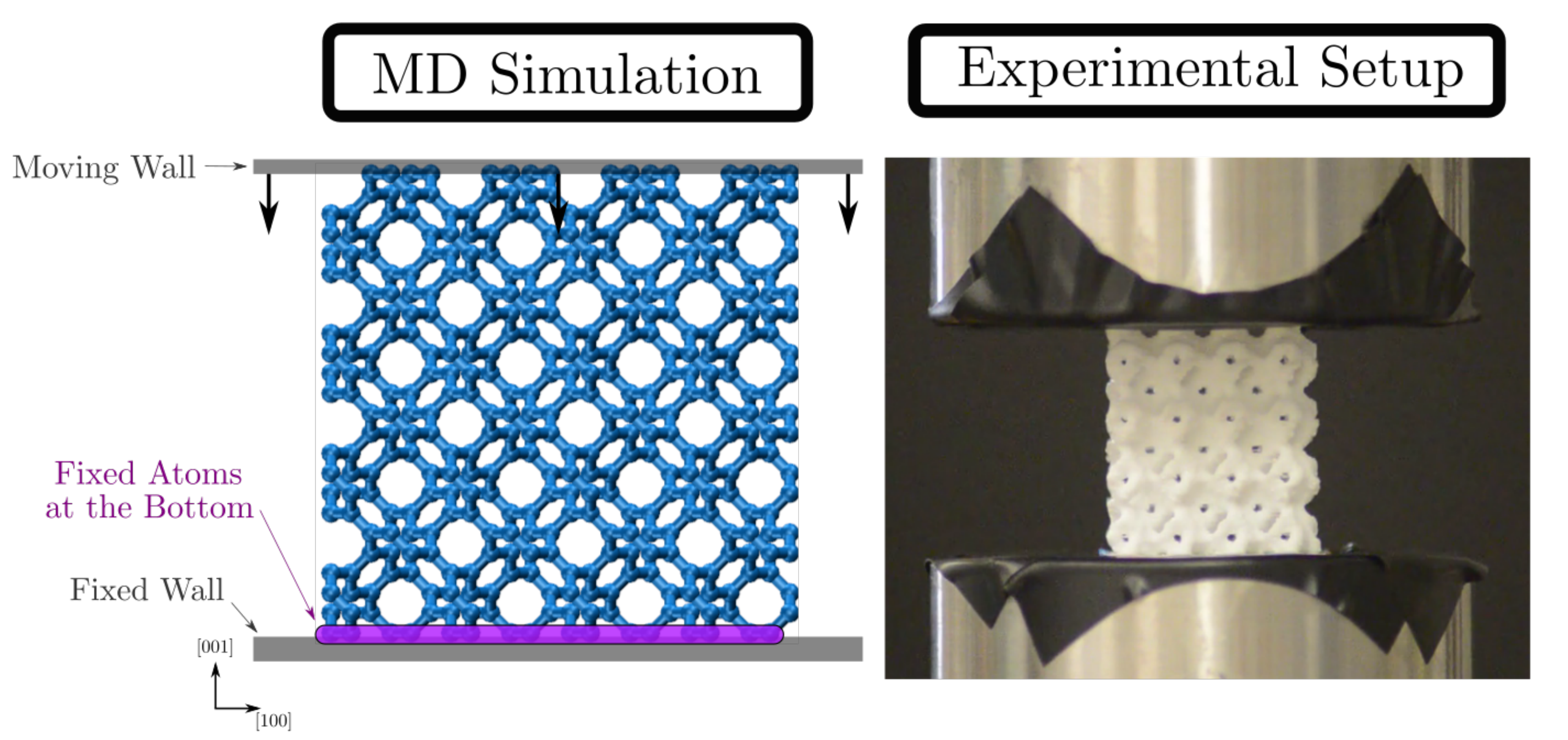}
    \caption{Scheme of the protocols used in MD simulations (left) and the experimental setup of the mechanical testings (right) of the 3D-printed structures.}
    \label{fig:setup}
\end{figure}

Fully-atomistic MD simulations were carried out using the Adaptive Interatomic Reactive Bond-Order (AIREBO)\cite{stuart_2000} potential as implemented by the Large-scale Atomic/Molecular Massively Parallel Simulator (LAMMPS)\cite{plimpton_1995}. For both D structures, the simulations were performed considering finite boundary conditions where the simulation box was set to be much larger than the structures to avoid suppression of lateral deformations that are consequence of the compression. Before mechanical tests, an NVT equilibration was carried out for $100$~ps at $10$~K (we used this low temperature to decrease thermal fluctuations in the stress-strain curves). Then, a uniaxial compression at a constant rate 10$^{-6}$~fs$^{-1}$ was performed also at $10$~K. As illustrated in Figure \ref{fig:setup}(a), a reflecting moving wall was used to compress the atomic structures along the [001]-direction, where some atoms at the bottom of the structures were kept fixed and a fixed reflecting wall was set below those fixed atoms.

To obtain the stress-strain curves, we calculated the compressive strain along the direction of loading, defined as
\begin{equation}
    \varepsilon_c = \frac{L_0 - L}{L_0},
\end{equation}
where $L_0$ is the initial length along [001] and $L$ is its current value during the compression. Furthermore, from the continuum interpretation of the virial stress \cite{subramaniyan_2008} we calculated the compressive stress, given by
\begin{equation}
    \sigma_{ij} = \frac{1}{V}\sum^N_{k=1} m_k v_{ki}v_{kj} + \frac{1}{V}\sum^N_{k=1}r_{ki}f_{kj}, \quad (i,j = x,y,z),
\end{equation}
where the indices $i,j = x,y,z$ correspond to the Cartesian directions, $m_k$ is the atomic mass of the $k$-th atom, $r_{ki}$ ($v_{ki}$) is the atomic position (velocity) of the $k$-th along the $i$ direction, $f_{kj}$ is the corresponding force on the $k$-th atom due to its neighbors and $V$ is the volume of the structure. The Young's modulus $E$ defined in the elastic regime is obtained by a linear fitting in the low-strain region (for more details, see the Supporting Information) where the stress-strain relationship can be approximated by
\begin{equation}
    \sigma_{zz} = E\varepsilon_c.
\end{equation}
In the same way, the Poisson's ration $\nu$ is obtained by a linear fitting for low values in the transverse vs longitudinal strain data (see the Supporting Information) as
\begin{equation}
    \nu = -\frac{\varepsilon_T}{\varepsilon_L},
\end{equation}
where $\varepsilon$ is the normal strain given by
\begin{equation}
    \varepsilon_L = \frac{L - L_0}{L_0},
\end{equation}
and $\varepsilon_T$ is the strain along any transverse direction ([100] or [010] in Figure \ref{fig:structures}).

The analysis of the local stress concentration was based on the values of the von Mises stress\cite{mises_1913} (which is helpful in the fracture analyses) per atom $k$ calculated by the relation
\begin{equation}
    \sigma^{k}_{v} = \sqrt{\frac{(\sigma^{k}_{xx} - \sigma^{k}_{yy})^2 + (\sigma^{k}_{yy} - \sigma^{k}_{zz})^2 + (\sigma^{k}_{xx} - \sigma^{k}_{zz})^2 + 6((\sigma^k_{xy})^2+(\sigma^k_{yz})^2+(\sigma^k_{zx})^2)}{2}}.
\end{equation}

\subsection{3D Printing}
The schwarzite optimized atomic structures from MD simulations were used as input to create macro models that are then 3D printed. The atomic positions were used to build interpolated surfaces like the ones shown on the left panel of Figure \ref{fig:surfaces} and corresponding 3D-printed structures are shown in the right panel. The 3D printed structures were obtained using a homemade CoreXY Fused Deposition Modelling 3D printer. It uses a commercial hot nozzle of 400 $\mu$m in diameter to extrude Poly Lactic Acid (PLA) polymer filaments of $1.75$~mm in diameters to print layers of 200~$\mu$m. The printing heads are moved along the XY-plane by two stepper motors following the H-frame type XY-positioning system\cite{itoh_2004,sollmann_2010} and the hotbed is moved along the z-axis by another couple of stepper motors. The printer control is based on an open-source Arduino microcontroller board Mega 2560 interfaced with a commercial RepRap Arduino Mega Pololu Shield (RAMPS).

In order to investigate the mechanical behavior of the 3D-printed samples (shown in the right panel of Figure \ref{fig:surfaces}), they were compressed on an MTS 793 testing machine. This testing machine consists of a fixed table and a hydraulic piston that is set to compress the sample at a constant rate of 5 mm/min up to 80\% strain. The hydraulic piston has a load cell to record the force during the test with a 200kN limit. To record the transverse dimensions changes during the compression tests a Nikon D7100 camera was used with 1280 pixels resolution and 23 frames/s pointed perpendicular to one side of the sample. With the frames obtained from the recorded video at constant time intervals, an intensity profile for each frame was plotted along the [100] and [001] axes. As the sample was white and the background green the channel blue of the frame provides a white sample and black background. Thus, in the intensity profile, the high values are attributed to the sample while the low values to the background. Then, by analyzing the intensity profile it was possible to see that the slopes of intensity were the sample borders. Measuring the distance between the slopes (in pixels) and applying a scale factor (pixels per millimeter taken from the first frame) it was possible to measure the face dimensions (in millimeters) and consequently their change values.

\subsection{Energy Absorption Properties}

Since schwarzites are porous (or cellular) materials, it is interesting to investigate their energy absorption characteristics\cite{gibson_1997}. The first one is the \textit{specific absorption energy} $W$ (energy per unit volume) given by the area under the stress-strain curve
\begin{equation}\label{eq:absorption-energy}
    W = \int_0^\varepsilon \sigma(\varepsilon')d\varepsilon',
\end{equation}
which is calculated up to a strain value $\varepsilon$. Another important energy-absorption parameter is the \textit{crush force efficiency}
\begin{equation}\label{eq:crush-force}
    \textnormal{CFE} = \frac{\sigma_{avg}(\varepsilon)}{\sigma_{max}(\varepsilon)},
\end{equation}
where $\sigma_{avg}(\varepsilon)$ and $\sigma_{max}(\varepsilon)$ are the average and maximum stress values up to the strain $\varepsilon$, respectively. An ideal energy absorber has $\textnormal{CFE}=1$.

Finally, the \textit{stroke efficiency} $S_E$, or \textit{densification strain} $\varepsilon_{dens}$, is the strain value that maximizes the \textit{energy absorption efficiency}
\begin{equation}
    \eta(\varepsilon) = \frac{1}{\sigma(\varepsilon)}\int_0^\varepsilon \sigma(\varepsilon')d\varepsilon'.
\end{equation}
For strain values greater than $S_E$, the material no longer behaves as an energy absorber. At this point, the structure enters in a regime of densification characterized by a rapid increase of the stress values on the stress-strain curve.

Previous works have already explored the energy absorption capabilities of TPMS-like materials\cite{wu_2015,pedrielli_2017,maskery_2017,sychov_2018,felix_2019}.

\section{Results and Discussions}

\begin{figure}[ht]
    \centering
    \includegraphics[scale=0.47]{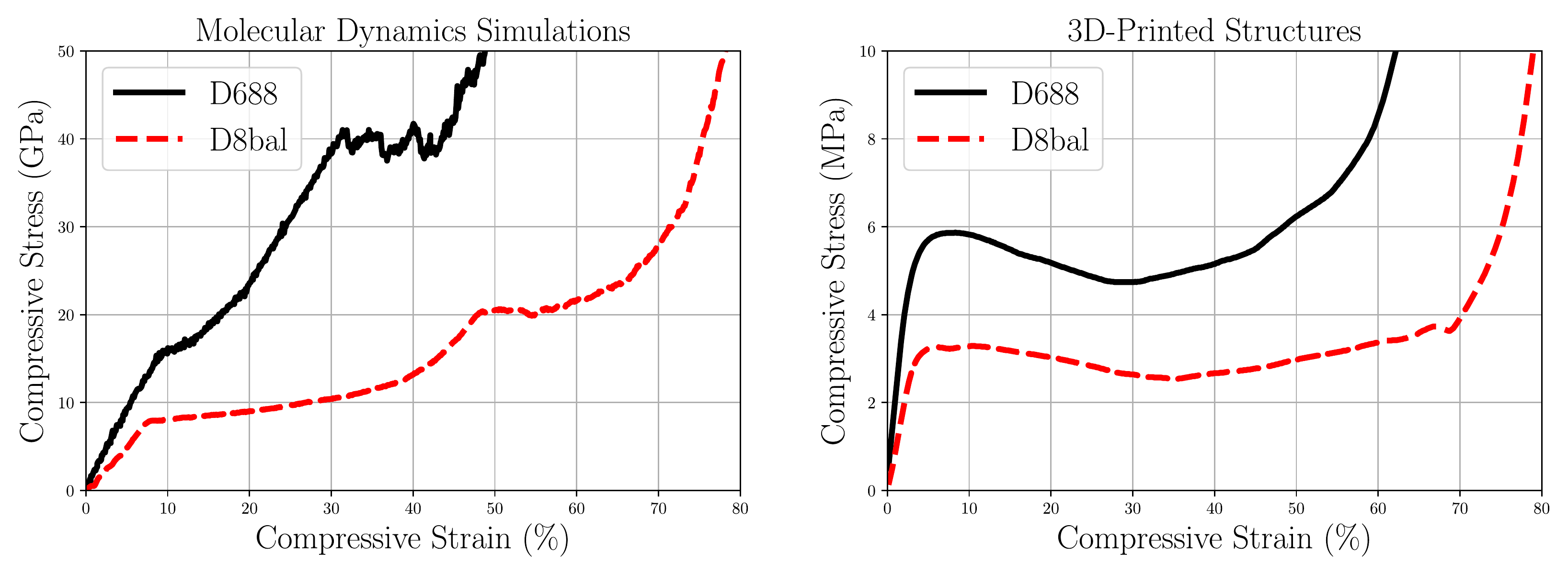}
    \caption{Stress-strain curves for: (a) MD simulations, and; (b) Mechanical testing of the 3D-printed structures.}
    \label{fig:stress-strain}
\end{figure}

In Figure \ref{fig:stress-strain} we present the stress-strain curves (uniaxial compression) for MD and 3D printed mechanical tests. For both cases, we can observe three regimes: (I) a linear behavior at low-strain values corresponding to the elastic regime; (II) a post-yielding collapse plateau due to plastic-like deformation, then followed by (III) a steep increase of stress values that characterizes the densification regime (for $\varepsilon>\varepsilon_{dens}$). These three regimes are characteristic of porous/cellular materials \cite{gibson_1997}.

From the linear regime of the stress-strain curves, we estimated Young's modulus and Poisson's ratio values (for details see Figures S1-S4 in the Supplementary Materials). The results are summarized in Table \ref{tab:elastic}. The Young's modulus values of atomic schwarzites are three orders of magnitude larger than that of the macroscopic 3D-printed structures. This is expected since in one case we have graphitic-like structures (carbon schwarzites) against the soft PLA-based 3D printed schwarzites. But even in this case, the D688$/$D8bal Young's modulus ratio values are close (0.69 and 0.51, from MD and 3D print results, respectively). On the other hand, Poisson's ratio values are expected to be mainly geometric-dependent (the trends should be scale and material independent), and they are indeed very close, as indicated in Table \ref{tab:elastic}. D8bal structures are softer (smaller values of $E$ compared to the D688), as their larger pore sizes result in higher "flatness" and/or lower density values. The smaller pore (atomic) sizes for D688 contributes to the suppression of the observed collapse plateaus (see Figure \ref{fig:stress-strain}). Another important observation is that the Young's modulus and Poisson's ratio values ordering are the same for MD and 3D print results, which indicate that some of the mechanical behavior is scale independent. This was also reported to other schwarzite families \cite{sajadi_2018,felix_2020}. 


\begin{table}[ht]
    \centering
    \begin{tabular}{|c|c|c|c|c|}
    \hline
        \textbf{Structure} & $\boldsymbol{E_{MD}}$ \textbf{[GPa]} & $\boldsymbol{E_{3D}}$ \textbf{[MPa]} & $\boldsymbol{\nu_{MD}}$ & $\boldsymbol{\nu_{3D}}$ \\
        \hline
        D688 & 177.55 & 189.31 & 0.31 & 0.27\\
        \hline
        D8bal & 123.39 & 97.11 & 0.05 & 0.04\\
         \hline
        D8bal/D688 & 0.69 & 0.51 & 0.16 & 0.15\\
        \hline
    \end{tabular}
    \caption{Elastic properties of D688 and D8bal schwarzites: Young's modulus for MD simulations ($E_{MD}$) and 3D-printed structures ($E_{3D}$) and their corresponding Poisson's ratio $\nu_{MD}$ and $\nu_{3D}$. The D8bal/D688 ratio values for these magnitudes are indicated in the last line of the table.}
    \label{tab:elastic}
\end{table}

The stress-strain behaviour observed in Figure \ref{fig:stress-strain} is typical of elastic-plastic foams\cite{gibson_1997}, where the plateau stress occurs due to the collapse/buckling of pores. This can be better evidenced in the Figure~\ref{fig:snapshots} and the videos 1 and 2 in the Supplementary Materials. The structures are remarkably resilient and can be deformed with elastic/plastic recovery more than 30-40\% in the atomic cases and more than 50\% for the 3D print ones. The compression leads to shear-like deformations that result in a parallelogram shape for large strain regimes, which can be clearly seen in Figure \ref{fig:stress-strain} for both atomistic and 3D printed structures. D8bal structures present a near-zero Poisson's ratio for low strain values (also listed in Table \ref{tab:elastic}) due to the re-entrant character of their pores. Such behavior is suppressed in the D688 structure as a consequence of their smaller pore sizes.

\begin{figure}[ht]
    \centering
    \includegraphics[scale=0.18]{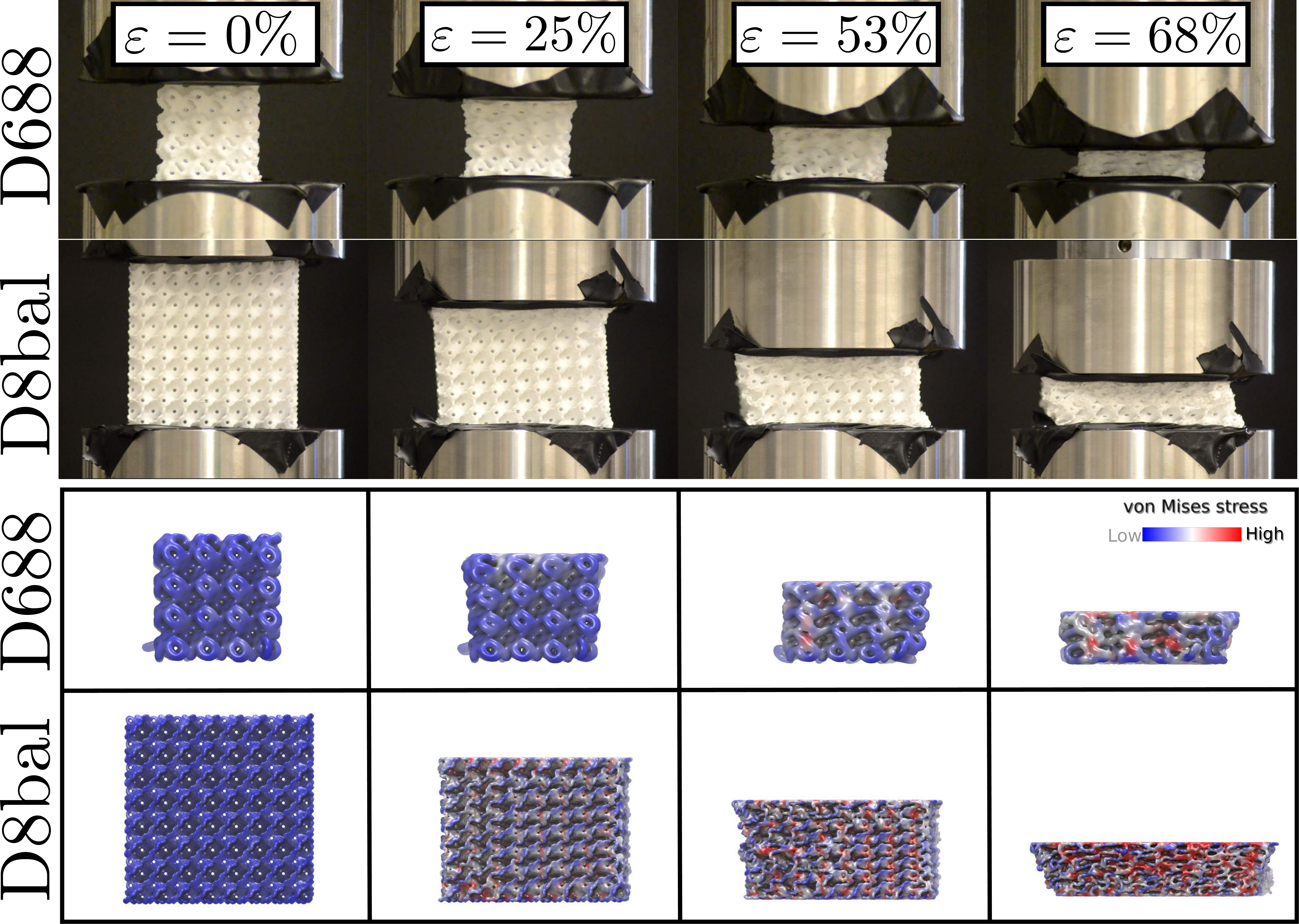}
    \caption{Representative snapshots from the mechanical compressive tests (top) and the von Mises stress values from MD simulations (bottom), at different strain stages.}
    \label{fig:snapshots}
\end{figure}

The energy absorption energy defined in Equation \ref{eq:absorption-energy} is usually calculated at $\varepsilon=S_E$ (stroke efficiency). However, in order to compare the two structures studied in this work we chose to evaluate $W$ at $\varepsilon=30 \%$, since at this value neither of the structures exhibit fracture. Also, the energy absorption efficiency values were obtained at $\varepsilon=S_E$. The results are summarized in Table \ref{tab:absorption}.

Similarly to the Young's modulus values, as expected, the absolute numbers for the MD and 3D $W$ are of different magnitudes. Again, when we compare the D8bal/D688 ratio values there is a good MD/3D print agreement. Similarly to the mechanical properties (discussed above), the significant differences between the D688 and D8bal can be again attributed to their differences in 'flatness', which helps to structurally stabilize the structures. 

These results for the absorption energy values of both D688 and D8bal are different from the ones recently reported \cite{sychov_2018} for 3D-printed PLA Diamond TPMS. This difference is possibly due to the fact that our 3D printed structures were obtained using a different approach than standard TPMS. Our 3D printed structures were obtained from models where the rendered surfaces are from energy minimized fully atomistic models, which provide a finer granularity (closer to the atomic models) than one obtained in the TPMS approach. This helps to decrease some structural instabilities that increases the stiffness of the structures as compared with pure TPMS \cite{miller_2016}.
The crush force efficiency values calculated here ($\sim 0.8$) are considerably higher than other structures investigated for crashworthiness applications reported in the literature\cite{martin_2010,tarlochan_2017}. These results suggest that the 3D printed schwarzites are good candidates for energy absorption applications.

\begin{table}[ht]
    \centering
    \begin{tabular}{|c|c|c|c|c|c|c|}
    \hline
        \textbf{Structure} & $\boldsymbol{W_{MD}}$ \textbf{[GJ/m$^3$]} & $\boldsymbol{W_{3D}}$ \textbf{[MJ/m$^3$]} & $\boldsymbol{S_{E,MD}}$ \textbf{[\%]} & $\boldsymbol{S_{E,3D}}$ \textbf{[\%]} & $\boldsymbol{\eta_{MD}}$ & $\boldsymbol{\eta_{3D}}$ \\
        \hline
        D688 & 5.92 & 1.53 & 31.02 & 54.13 & 0.16 & 0.42 \\
        \hline
        D8bal & 2.30 & 0.87 & 38.95 & 68.62 & 0.26 & 0.56  \\
        \hline
        D8bal/D688 & 0.39 & 0.57 & 1.26 & 1.27 & 1.63 & 1.33\\
         \hline
    \end{tabular}
    \caption{Energy absorption characteristics of D688 and D8bal schwarzites. The values of the specific absorption energy $W$, stroke efficiency $S$, and the energy absorption efficiency $\eta$ calculated at $\varepsilon=S_E$ (stroke efficiency). MD and 3D refer to the molecular dynamics and 3D print results, respectively. The D8bal/D688 ratio values for these magnitudes are indicated in the last line of the table.}
    \label{tab:absorption}
\end{table}

\section{Conclusions}

We investigated the mechanical behavior (deformations under compressive strains) and energy absorption properties of the D688 and D8bal diamond schwarzites. We carried out fully-atomistic reactive molecular dynamics (MD) simulations. The optimized atomic models were used to generate macro models that were 3D printed (thermoplastic PolyLactic Acid (PLA) polymer filaments). The stress-strain curves of the atomistic and 3D printed (3D) structures exhibit three regimes: a linear, a post-yielding collapse plateau due to plastic-like deformation, then followed by a steep increase of stress values that characterizes the densification regime. The estimated Young's modulus values were 177.55 GPa/189.31 MPa and 123.39 GPa/97.11 MPa for the MD/3D D688 and D8bal, respectively. D8bal presents a near-zero Poisson's ratio for both MD and 3D structures. Interestingly, some aspects of the mechanical behavior may be scale independent. A layer-by-layer-like (pore closing/collapse) deformation mechanism (both structures start to deform from the top) was observed for atomistic and 3D-printed structures. Also, the structures exhibit high energy absorption per unit volume. The estimated specific absorption energy values were (5.92 GJ/m$^3$)/(1.53 MJ/m$^3$) and (2.30 GJ/m$^3$)/(0.87 MJ/m$^3$) for D688 and D8bal MD/3D, respectively. For both atomic and macroscopic schwarzites, D8bal energy absorption efficiency ($\eta$) is considerably higher than the values for D688 (1.63 and 1.33 for MD and 3D, respectively). The estimated crush force efficiency was of the order of $\sim 0.8$. These findings suggest that these 3D printed diamond schwarzites are good candidates for energy absorption applications.

\begin{acknowledgement}

This work was financed in part by the Coordenacão de Aperfeiçoamento de Pessoal de Nível Superior - Brasil (CAPES) - Finance Code 001 and CNPq and FAPESP. The authors thank the Center for Computational Engineering \& Sciences (CCES) at Unicamp for financial support through the FAPESP/CEPID Grant 2013/08293-7 and LNNano/CNPEM for providing the testing equipment. They also thank Osmar Roberto Bagnato and Francini Hesse for useful discussions and support given in some experiments.

\end{acknowledgement}


\bibliography{references}

\end{document}